\documentclass{emulateapj}
\usepackage{apjfonts}
\usepackage{hyperref}

\bibpunct{(}{)}{;}{a}{}{,}

\newcommand{\vhigh}{\vphantom{\frac{1}{1}}}
\newcommand{\vhighs}{\vphantom{1^1_1}}
\newcommand{\ra}{r_\mathrm{+}}
\newcommand{\rp}{r_\mathrm{-}}
\newcommand{\calE}{E}
\newcommand{\pseudo}{Q}

\newcommand{\txa}{{\mathrm{a}}}
\newcommand{\txc}{{\mathrm{c}}}
\newcommand{\txd}{{\mathrm{d}}}
\newcommand{\txe}{{\mathrm{e}}}
\newcommand{\txs}{{\mathrm{s}}}
\newcommand{\tqp}{\mathrm{qp}}
\newcommand{\tob}{\mathrm{obs}}
\newcommand{\tlib}{\mathrm{lib}}
\newcommand{\tmin}{\mathrm{min}}
\renewcommand{\leq}{\leqslant}
\renewcommand{\geq}{\geqslant}
\newcommand{\Laplace}{\mathop{\mathcal{L}}}
\newcommand{\Mellin}{\mathop{\mathcal{M}}}
\newenvironment{Hmatrix}{\begin{array}{r@{\,,\,}l}}{\end{array}}

\shorttitle{The structure of DM halos with universal properties}
\shortauthors{E.~Van Hese, M.~Baes \& H.~Dejonghe}

\begin{document}

\title{The dynamical structure of dark matter halos with universal properties}

\author{Emmanuel Van Hese, Maarten Baes and Herwig Dejonghe}
\affil{Sterrenkundig Observatorium, Universiteit Gent, Krijgslaan 281 S9, B-9000 Gent, Belgium}
\email{emmanuel@heze.ugent.be}
\email{maarten.baes@ugent.be}
\email{herwig.dejonghe@ugent.be}

\begin{abstract}
  $N$-body simulations have unveiled several apparently universal
  properties of dark matter halos, including a cusped density profile,
  a power-law pseudo phase-space density $\rho/\sigma_r^3$, and a
  linear $\beta-\gamma$ relation between the density slope and the
  velocity anisotropy. We present a family of self-consistent
  phase-space distribution functions $F(E,L)$, based on the
  Dehnen-McLaughlin Jeans models, that incorporate these universal
  properties very accurately. These distribution functions, derived
  using a quadratic programming technique, are analytical, positive
  and smooth over the entire phase space and are able to generate
  four-parameter velocity anisotropy profiles $\beta(r)$ with
  arbitrary asymptotic values $\beta_0$ and $\beta_\infty$. We discuss
  the orbital structure of six radially anisotropic systems in detail
  and argue that, apart from its use for generating initial conditions
  for $N$-body studies, our dynamical modeling provides a valuable
  complementary approach to understand the processes involved in the
  formation of dark matter halos.
\end{abstract}


\section{Introduction}
\label{introduction.sec}
\defcitealias{2007A&A...471..419B}{Paper~I}

In theoretical astrophysics the steady increase in computational power
has sparked a proportional interest and progress in the study of
large-scale structure formation. In particular, as $N$-body
simulations of cold dark matter halos have become more detailed,
several ''universal'' properties have emerged. We highlight three
important characteristics.

Firstly, numerous cosmological studies
\citep[e.g.][]{1991ApJ...378..496D, 1994ApJ...434..402C,
1996ApJ...462..563N, 1997ApJ...477L...9F, 1997ApJ...490..493N,
1997ApJ...485L..13C, 1998ApJ...499L...5M, 1999MNRAS.310.1147M,
2000ApJ...529L..69J} revealed similar density profiles over several
orders of magnitude in halo mass, with a central cusp and an a
$\rho(r) \propto r^{-3}$ falloff at large radii. These generalized NFW
models can be described by a general 3-parameter family, referred to
as the Zhao models (or $\alpha\beta\gamma$-models)
\citep{1990ApJ...356..359H,1996MNRAS.278..488Z}.  They are defined by
the density
\begin{equation}\label{zhao}
\rho(r) =
\frac{2^{(\gamma_\infty-\gamma_0)/\eta}\,\rho_\txs}{\left(r/r_\txs\right)^{\gamma_0} 
\left(1 + \left(r/r_\txs\right)^\eta\right)^{(\gamma_\infty-\gamma_0)/\eta}},
\end{equation}
or in terms of the logarithmic slope,
\begin{equation}
\gamma(r) = -\frac{\txd\ln\rho}{\txd\ln r}(r) 
= \frac{\gamma_0 + \gamma_\infty \left(r/r_\txs\right)^\eta}{1 + \left(r/r_\txs\right)^\eta}.
\end{equation}
Although in recent years alternative profiles based on the S\'ersic
law have produced equally good results \citep{2004MNRAS.349.1039N,
2005ApJ...624L..85M}, the generalized NFW models remain very popular
and successful to represent dark matter halos.

A second relation was found by \citet{2001ApJ...563..483T}. These
authors identified that the quantity $\pseudo(r)=\rho/\sigma^3(r)$,
which has become known as the pseudo phase-space density, behaves as a
power law over 2-3 orders of magnitude in radius inside the virial
radius,
\begin{equation}
\pseudo(r) \propto r^{-\alpha}.
\end{equation}
Other studies \citep[e.g.][]{2004MNRAS.351..237R,2004MNRAS.352.1109A}
have confirmed the scale-free nature of $\pseudo(r)$, and their
results indicate that its slope lies in the range $\alpha=1.90\pm0.05$.
This property is remarkable since the density $\rho(r)$ nor the
velocity dispersion $\sigma(r)$ separately show a power-law behavior.

Finally, the velocity anisotropy profiles $\beta(r)$ of dark matter
systems also evolve toward a similar shape, steepening gradually from
isotropic in the center to radially anisotropic in the outer
regions. \citet{2006NewA...11..333H} suggest a nearly linear relation
between the logarithmic density slope $\gamma(r)$ and the velocity
anisotropy profile, based on various types of equilibrated
simulations. They proposed the $\beta-\gamma$ relation
\begin{equation}
  \beta(\gamma) \simeq 1 - 1.15(1 + \gamma/6).
\end{equation}

Several theoretical studies have been made to investigate whether
solutions of the Jeans equation exist that encompass the observed
properties of dark matter halos. In particular,
\citet{2005MNRAS.363.1057D} investigated the anisotropic Jeans
equation constrained by a slightly different form of the pseudo
phase-space density, namely $\pseudo_r(r)=\rho/\sigma_r^3$ with
$\sigma_r(r)$ the radial velocity dispersion.  They found a special
solution, namely an analytical self-consistent potential-density pair
of the form (\ref{zhao}) with an anisotropy profile
\begin{equation}
  \beta(r)
  =
  \frac{\beta_0+\beta_\infty(r/r_\txa)^{2\delta}}{1+(r/r_\txa)^{2\delta}},
\label{betagen}
\end{equation}
that also has an exactly linear $\beta-\gamma$ relation (in other
words, $r_\txa=r_\txs$ and $2\delta = \eta$).  As these Jeans models
satisfy the three universal relations mentioned above, we can consider
them as representative models for realistic dark matter halos.

The goal of this paper is to take the analytical study of dark matter
systems a step further, i.e.\ to look for full dynamical models that
encompass the universal properties found in $N$-body simulations. In
concreto we will look for phase-space distribution functions (DFs)
$F(\vec{r},\vec{v})$ that self-consistently generate the required
density, potential, and anisotropy profiles encountered in dark matter
halos. Such dynamical models would provide a very useful
complementary approach to gain insight into the structure of dark
matter halos.

We shall focus on spherical models, for which a dynamical description
simplifies substantially as in this case the DF can be expressed as a
function $F(\calE,L)$ of the binding energy and the angular
momentum. But even if we limit ourselves to the spherical case, it is
not straightforward to obtain such dynamical models.  A simple
approach is to solve the Jeans equation and approximate the velocity
dispersion profiles by a multivariate Gaussian
\citep{1993ApJS...86..389H}. \citet{2004ApJ...601...37K} demonstrated
however that these systems are far from equilibrium, thus losing their
initial dynamical structure as they evolve to non-Gaussian velocity
distributions in subsequent $N$-body simulations. We therefore need
the tools to derive full self-consistent equilibrium models, governed
by DFs with a sufficiently general velocity anisotropy profile.

Simple analytical models are found for only a few special cases, such
as the Plummer, Hernquist, isochrone, and $\gamma$-models
\citep[e.g.][]{1987MNRAS.224...13D, 1990ApJ...356..359H,
2002A&A...393..485B, 2006AJ....131..782A, 2007MNRAS.375..773B}, none
of which are able to describe realistic dark matter halos.  For
general potential-density pairs, \citet{1979PAZh....5...77O} and
\citet{1985AJ.....90.1027M} provided an algorithm to construct
dynamical models with a specific velocity anisotropy profile of the
form (\ref{betagen}), with $\beta_0=0$, $\beta_\infty=1$ and
$\delta=1$, leaving $r_\txa$ as a free parameter. While the
Osipkov-Merritt method has been widely adopted
\citep[e.g.][]{2000ApJS..131...39W, 2001MNRAS.321..155L,
2004ApJ...601...37K}, comparison with simulations shows that such
anisotropy profiles are too steep to describe dark matter
\citep{2005MNRAS.363..705M}. Halos are not completely radial at
infinity (i.e.\ $\beta_\infty<1$) and the linear $\beta-\gamma$
relation can only be realized with a lower transition rate $\delta<1$.
On a dynamical note, the associated DFs are of the form $F(Q)$ with
$Q=\calE - L^2/2r_\txa^2$, creating an unphysical cut-off boundary for
orbits with $Q<0$. Hence, the Osipkov-Merritt framework is too limited
to generate realistic dark matter systems, and a more extensive method
is needed.

Recently, \citet{2008MNRAS.tmp..719W} presented an interesting
approach to generate dynamical models for potential-density
pairs with a more general anisotropy profile.  They proposed to
express the DF as a separable function of the form
$f_\calE(\calE)\,f_L(L)$ where $f_L(L)$ is a double power-law function
with three parameters $\beta_0$, $\beta_\infty$, and $L_0$. Once their
values have been determined, the function $f_\calE(\calE)$ is derived
from the observed density profile by a numerical inversion. This
technique yields a three-parameter anisotropy profile that resembles
eq.~(\ref{betagen}), where $L_0$ has a similar role as $r_\txa$, and a
fixed transition rate $0.5<\delta<1$. In this manner, the authors were
able to construct models with an NFW density with velocity dispersion
profiles that agreed with their dark matter simulations. 

In this paper, we present a technique that enables us to obtain
dynamical models with exactly the four-parameter anisotropy
profile~(\ref{betagen}).  We demonstrated in a previous paper
\citep[][hereafter
\citetalias{2007A&A...471..419B}]{2007A&A...471..419B} how this can be
achieved, by postulating a separable parameterized form of the
so-called augmented density, which provides an equivalent description
of a dynamical system. In certain special cases the transformation
from the augmented density to the DF is analytically tractable. We
were thus able to derive a family of DFs for the generalized Plummer
models (also called $\alpha$-models or Veltmann models,
\citet{1979AZh....56..976V}) with a linear $\beta-\gamma$ relation.

Since the generalized Plummer potential-density pairs form a special
class of the Zhao models~(\ref{zhao}), we will now demonstrate that
this result is a first step toward more representative dark matter
profiles.  In particular, we seek a family of DFs for the
Dehnen-McLaughlin systems, because of their unique property to satisfy
a universal density, a power law $\pseudo_r(r)$, and a linear
$\beta-\gamma$ relation. We use a quadratic programming technique
\citep{1989ApJ...343..113D} to build DFs as a linear combination of
base functions of the form derived in
\citetalias{2007A&A...471..419B}.  In this manner, we demonstrate that
it is indeed possible to generate full dynamical models for the
Dehnen-McLaughlin Jeans models, thus providing DFs that encompass the
observed properties of dark matter halos.

Our paper is organized as follows. In Section~\ref{preliminaries.sec}
we describe the notion of a dynamical model and we briefly summarize
the main aspects of the Dehnen-McLaughlin halos.  Next we outline our
modeling technique in Section~\ref{modeling.sec}: we explain the
quadratic programming algorithm and we recapitulate the functions that
we derived in \citetalias{2007A&A...471..419B}. With these components,
we can build a library to apply the QP-method to the Dehnen-McLaughlin
halos. In Section~\ref{results.sec} we present our results for a set
of systems with different velocity anisotropy profiles, and we discuss
the moments, phase-space DFs, and energy and angular momentum
distributions for these models. Finally, we formulate our conclusions
in Section~\ref{conclusions.sec}.

\section{Preliminaries}
\label{preliminaries.sec}

\subsection{Dynamical models}
\label{dynmodels.sec}

The dynamical structure of a gravitational equilibrium system is
completely determined by the DF $F(\vec{r},\vec{v})$, which describes
the probability distribution of particles in six-dimensional phase
space. A dynamical model is only physical if its DF is non-negative
everywhere.  In the case of spherical symmetry, this DF
can be written as a function $F(\calE,L)$ of two isolating integrals,
namely the binding energy and the angular momentum
\begin{eqnarray}
  \calE 
  &=&
  \psi(r) - \frac{1}{2}\,v_r^2 - \frac{1}{2}\,v_T^2,
  \\
  L &=& r\,v_T,
\end{eqnarray}
with 
\begin{equation}
  v_T = \sqrt{v_\theta^2+v_\varphi^2},
\end{equation}
the transverse velocity, and $\psi(r)$ the positive binding potential.
For circular orbits the integrals of motion can be written in function
of the radius $r$,
\begin{eqnarray}
  \calE_\txc(r)
  &=&
  \psi(r) + \frac{r}{2}\frac{\txd \psi}{\txd r}(r),
  \\
  L_\txc(r) &=& -r^3 \frac{\txd \psi}{\txd r}(r),
  \label{circorbs}
\end{eqnarray}
which can be solved to obtain $\calE_\txc(L)$ and $L_\txc(\calE)$.
All dynamical properties can be derived from the DF, such as the
anisotropic velocity moments
\begin{equation}
  \mu_{2n,2m}(r) 
  = 
  2\pi M_\mathrm{tot} \int_{-\infty}^{+\infty}\txd v_r \int_0^{+\infty}
  F(\calE,L)\,v_r^{2n}\,v_T^{2m+1}\,\,\txd v_T,
\label{momdf}
\end{equation}
with $M_\mathrm{tot}$ the total mass.  If the system is
self-consistent, then the density $\rho(r)= \mu_{00}(r)$ is connected
to the potential via the Poisson equation
\begin{equation}
  \frac{1}{r^2}\,\frac{\txd}{\txd r}
  \left(r^2\,\frac{\txd\psi}{\txd r}\right)(r)
  =
  -4\pi G\rho(r).
\end{equation}
Furthermore, the second-order moments determine the radial and
tangential velocity dispersions $\mu_{20}(r)=\rho\sigma_r^2(r)$ and
$\mu_{02}(r)=2\rho\sigma_\theta^2(r)$, and the velocity anisotropy
profile
\begin{equation}
  \beta(r) 
  =
  1 - \frac{\sigma_\theta^2(r)}{\sigma_r^2(r)}.
\end{equation}
A DF is only fully determined when all velocity moments~(\ref{momdf})
are known \citep{1987MNRAS.224...13D}. However, it is already far from
trivial to obtain \emph{any} non-negative DF that adequately generates
a given density $\rho(r)$ and anisotropy profile $\beta(r)$. In this
paper, we will focus our attention to one specific family of halos,
namely the Dehnen-McLaughlin Jeans models, and we demonstrate that a family
of DFs can indeed be constructed for these systems.

\subsection{The Dehnen-McLaughlin halos}
\label{dehnen.sec}

In the context of theoretical dark matter studies, the model derived by
\citet{2005MNRAS.363.1057D} is of particular interest. We summarize
their main results in this section. Instead of fitting a parameterized
density profile to $N$-body simulations, they investigated the
solution space of the Jeans equation to search for models that
explicitly obey the power-law behavior of the pseudo phase-space
density. With the extra condition of a linear $\beta-\gamma$ relation
they found a critical solution that satisfies the condition
\begin{equation}
  \frac{\rho}{\sigma_r^\varepsilon}(r) = \frac{\rho}{\sigma_r^\varepsilon}(r_\txs)\,
  \left(\frac{r}{r_\txs}\right)^{-\alpha_\mathrm{crit}},
  \label{pseudodens}
\end{equation}
with $r_\txs$ a scale radius.  In the remainder of this paper, we
adopt the common value $\varepsilon=3$ and use the notation
$\pseudo_r(r)=\rho/\sigma_r^3$. Dehnen \& McLaughlin derived for this
case the exponent
\begin{eqnarray}
  \alpha_\mathrm{crit} &=& \eta+\frac{3}{2},\\
  \eta &=& \frac{4-2\beta_0}{9},
  \label{etadehnen}
\end{eqnarray}
with $\beta_0$ the central velocity anisotropy.  The corresponding
potential-density pair is fully analytical, given by
\begin{eqnarray}
  \psi(r) &=& \frac{GM_{\mathrm{tot}}}{r_\txs}\,\frac{1}{\eta}\,
  B_{\frac{1}{1+x^\eta}}
  \left( \frac{1}{\eta},\frac{1-\beta_0}{\eta}+\frac{1}{2} \right),
\label{psidehnen}
\\
  \rho(r) &=& \frac{4+\eta-2\beta_0}{8\pi}\,\frac{M_{\mathrm{tot}}}{r_\txs^3}\,
  x^{-\gamma_0}\,\left(1+x^\eta\right)^{-(\gamma_\infty-\gamma_0)/\eta},
\label{rhodehnen}
\end{eqnarray}
where $x = r/r_\txs$, $B_y(a,b)$ is the incomplete beta function and
\begin{eqnarray}
  \gamma_0 &=& \frac{7+10\beta_0}{9},\\
\label{gammadehnen0}
  \gamma_\infty &=& \frac{31-2\beta_0}{9}.
\label{gammadehneni}
\end{eqnarray}
The density can be equivalently written in terms of the slope
$\gamma(r)$, which has the same elegant form as the velocity
anisotropy profile $\beta(r)$
\begin{eqnarray}
  \gamma(r) &=& \frac{\gamma_0+\gamma_\infty x^\eta}{1+x^\eta},
  \\
  \beta(r) &=& \frac{\beta_0+\beta_\infty x^\eta}{1+x^\eta}.
\label{betadehnen}
\end{eqnarray}
Finally, the authors derived the corresponding velocity dispersions
\begin{eqnarray}
  \sigma_r^2(r) &=& \frac{1}{4+\eta-2\beta_\infty}\,
  \frac{GM_{\mathrm{tot}}}{r_\txs}\,x^{-1}\,
  \left(\frac{x^\eta}{1+x^\eta}\right)^{(\gamma_\infty-\gamma_0)/\eta-2},
\label{sigmardehnen}
  \\
  \sigma_\theta^2(r) &=& \sigma_\varphi^2(r) = \frac{1}{2}\sigma_T^2(r) =
  \left(1-\beta(r)\vhighs\right)\,\sigma_r^2(r).
\label{sigmatdehnen}
\end{eqnarray}
Eqs.~(\ref{pseudodens})-(\ref{sigmatdehnen}) characterize the
Dehnen-McLaughlin models; we refer to their paper for more details.
These systems are determined by five parameters: the exponent
$\varepsilon$ in the pseudo phase-space density, two
scaling constants i.e.\ the total mass $M_{\mathrm{tot}}$ and the
scale radius $r_\txs$, and the asymptotic anisotropy parameters
$\beta_0$ and $\beta_\infty$. The authors also noted the remarkable
property that the shape of the density profile (and hence the
gravitational potential) only depends on $\beta_0$ and not on
$\beta_\infty$.

While the Dehnen-McLaughlin Jeans models are derived from theoretical
considerations, they also fit adequately galaxy-sized and certain
cluster-sized halos generated by $N$-body simulations
\citep{2005MNRAS.364..665D,2006AJ....132.2685M}. But as mentioned in
the previous Section, the density and velocity dispersions alone do
not determine the complete dynamical state of dark matter systems. We
therefore aim to incorporate these profiles into self-consistent
dynamical models, described by non-negative DFs.

\section{The modeling technique}
\label{modeling.sec}

To find DFs that describe the Dehnen-McLaughlin Jeans models, we adopt
the mathematical framework that we derived in
\citetalias{2007A&A...471..419B}. These tools enabled us to construct
a family of components with the anisotropy profile~(\ref{betagen}), by
means of the powerful augmented density concept. We summarize these
results, and we demonstrate how to build a linear combination of these
components using a quadratic programming (QP) technique
\citep{1989ApJ...343..113D}. In this manner, we can fit a dynamical
model to a given halo.

\subsection{The augmented density concept}

Since our dynamical models have to reproduce the
moments~(\ref{rhodehnen}), (\ref{sigmardehnen}) and
(\ref{sigmatdehnen}), we first seek DFs that are specifically designed
for this task. This can be done by introducing the augmented densities
$\tilde\rho(\psi,r)$ \citep{1986PhR...133..217D}, which extend the
densities to explicit functions of both the radius and the
gravitational potential. Like the DF, the augmented density 
uniquely determines the dynamical state of a spherical equilibrium
system. Both functions are connected by the relation
\begin{equation}
  \tilde\rho(\psi,r)
  = 2\pi M_\mathrm{tot} \int_0^{\psi}\txd \calE\!
  \int_0^{2(\psi-\calE)}\!\!\!
  \frac{F(\calE,r\,v_T)}{\sqrt{2(\psi-\calE) - v^2_T}}\,\txd v^2_T.
\label{augdens}
\end{equation}
If we define that all functions are zero when their arguments lie
outside their physical bounds, we can use the Laplace-Mellin
transforms
\begin{eqnarray}
  \Laplace_{\calE\rightarrow\xi}\,\Mellin_{L\rightarrow\lambda}
  \,\left\{F\right\}
  &=& 
  \int_0^{+\infty}\!\!\!\txe^{-\xi\calE}\,\txd\calE \,
  \int_0^{+\infty}\!\!\!L^{\lambda-1}F(\calE,L)\,\txd L,\\
  \Laplace_{\psi\rightarrow\xi}\,\Mellin_{r\rightarrow\lambda}
  \,\left\{\tilde\rho\right\}
  &=& 
  \int_0^{+\infty}\!\!\!\txe^{-\xi\psi}\,\txd\psi \,
  \int_0^{+\infty}\!\!\!r^{\lambda-1}\tilde\rho(\psi,r)\,\txd r,
\label{laplacemellin}
\end{eqnarray}
with the inverse transforms
\begin{eqnarray}
  F(\calE,L)
  &=& 
  \frac{-1}{4\pi^2}\!\!\!
  \int_{\xi_0-i\infty}^{\xi_0+i\infty}\!\!\!\txe^{\xi\calE}\,\txd\calE \,
  \!\!\!\int_{\lambda_0-i\infty}^{\lambda_0+i\infty}
  \!\!\!\!\!L^{-\lambda}\!\!\!
  \Laplace_{\calE\rightarrow\xi}\,\Mellin_{L\rightarrow\lambda}\!\!
  \left\{F\right\}\,\txd L,\\
  \tilde\rho(\psi,r)
  &=& 
  \frac{-1}{4\pi^2}\!\!\!
  \int_{\xi_0-i\infty}^{\xi_0+i\infty}\!\!\!\txe^{\xi\psi}\,\txd\psi \,
  \!\!\!\int_{\lambda_0-i\infty}^{\lambda_0+i\infty}
  \!\!\!\!\!r^{-\lambda}\!\!\!
  \Laplace_{\psi\rightarrow\xi}\,\Mellin_{r\rightarrow\lambda}\!\!
  \left\{\tilde\rho\right\}\,\txd r.
\label{inverselaplacemellin}
\end{eqnarray} 
Combining these expressions with (\ref{augdens}), we obtain
\begin{eqnarray}
  \Laplace_{\psi\rightarrow\xi}\,\Mellin_{r\rightarrow\lambda}
  \,\left\{\tilde\rho\right\}
  &=& 
  2\pi M_\mathrm{tot} 
  \int_0^{+\infty}\!\!\!\txe^{-\xi\psi}\,\txd\psi
  \int_0^{+\infty}\!\!\!v_T^{-\lambda}L^{\lambda-1}\,\txd L \nonumber\\
  &\times&
  \int_0^{\psi}\!\!\txd \calE\!
  \int_0^{2(\psi - \calE)}\!\!\!
  \frac{F(\calE,L)\,\txd v^2_T}{\sqrt{2\left(\psi-\calE \right) - v^2_T}}.
\end{eqnarray}
Interchanging the integrals yields the formal relation
\begin{equation}
  \Laplace_{\calE\rightarrow\xi}\,\Mellin_{L\rightarrow\lambda}
  \,\left\{F\right\}
  = 
  \frac{2^{\lambda/2}}{M_\mathrm{tot}(2\pi)^{3/2}}
  \frac{\xi^{(3-\lambda)/2}}{\Gamma\left(1 - \frac{\lambda}{2}\right)}
  \,\Laplace_{\psi\rightarrow\xi}\,\Mellin_{r\rightarrow\lambda}
  \,\left\{\tilde\rho\right\}.
  \label{rhodf}
\end{equation}
The main strength of the augmented density formalism is the ability to
impose very specific conditions on these functions, to obtain our
objective. More precisely, we demonstrated in
\citetalias{2007A&A...471..419B} that separable functions of the form
\begin{equation}
  \tilde\rho(\psi,r)
  =
  f(\psi)\,g(r),
\end{equation}
are particularly interesting, since in this case $g(r)$
is directly related to the velocity anisotropy profile,
\begin{equation}
  \beta(r)
  =
  -\frac{1}{2}\frac{\txd\ln g}{\txd\ln r}(r).
  \label{betadv}
\end{equation}
With this formalism it is therefore at least formally possible
to generate a DF for a dynamical system with a given potential
$\psi(r)$, density $\rho(r)$, and velocity anisotropy profile
$\beta(r)$. If we postulate an anisotropy profile of the form
(\ref{betagen}), the modeling procedure reduces to the
computation of $f(\psi)$ such that
\begin{equation}
  \rho(r) 
  = 
  f\left(\psi(r)\vhighs\right)\,
  \left(\frac{r}{r_\txa}\right)^{-2\beta_0}
  \left(1+\frac{r^{2\delta}}{r_\txa^{2\delta}}\right)^{\beta_\delta},
\label{rhoisrho}
\end{equation}
with $0<\delta\leq 1$ and
\begin{equation}
  \beta_\delta
  =
  \frac{\beta_0-\beta_\infty}{\delta}.
\end{equation}
Evidently, the exact solution $f(\psi)$ is generally of numerical
form, except for a few special cases. This creates a problem to
recover the DF from the augmented density, because a direct inversion
of the Laplace-Mellin transformation in (\ref{rhodf}) is in general
numerically unstable \citep{1986PhR...133..217D}. Indeed, the double
integration in (\ref{augdens}) smooths features in $F(\calE,L)$ and
the inversion procedure of determining $F(\calE,L)$ from
$\tilde\rho(\psi,r)$ has the delicate job of unsmoothing the
information contained in $\tilde\rho(\psi,r)$.  Consequently, a direct
inversion can only be performed safely for sufficiently simple forms
of $f(\psi)$.

We therefore take an alternative approach. Instead of solving
(\ref{rhoisrho}) directly, we will approximate the given density
profile by a linear combination of simple base functions
$\tilde\rho_i(\psi(r),r)$ for which the Laplace-Mellin inversions are
analytical. The corresponding DF is then simply the same linear
combination of the associated base DFs $F_i(\calE,L)$, resulting in an
analytically tractable function. This can be achieved by a least
squares fit to a set of density data points, defining a quadratic
programming problem in the unknown coefficients. Various authors have
successfully used a similar modeling technique
\citep{1993MNRAS.264..712K,1994ApJ...432..575M,1998MNRAS.295..197G}.
A QP-algorithm, suited for this task, has been developed in our
department \citep{1989ApJ...343..113D}, which we outline in the
following Section.

\subsection{The quadratic programming procedure}
\label{qp.sec}

Consider a given potential $\psi(r)$, an anisotropy profile $\beta(r)$
of the form (\ref{betagen}) and a set of $M$ density data points
$\rho_\tob(r_m)$, $m=1,\ldots,M$. To model these data, we first
construct a library of $N_\tlib$ base functions 
\begin{equation}
  \tilde\rho_i(\psi,r)
  =
  f_i(\psi)\,
  \left(\frac{r}{r_\txa}\right)^{-2\beta_0}
  \left(1+\frac{r^{2\delta}}{r_\txa^{2\delta}}\right)^{\beta_\delta}, 
\end{equation}
with $f_i(\psi)$ sufficiently simple to compute the associated
distributions $F_i(\calE,L)$. From these components, we can extract
the corresponding values
\begin{equation}
  \rho_i(r_m) 
  = 
  \tilde\rho_i\left(\psi(r_m),r_m\vhighs\right),\qquad m=1,\ldots,M.
\end{equation}
Our aim is now to construct a linear combination of $N$ components
from this library that provides an adequate fit to the given
data. Such a fit can be obtained in a statistically meaningful way by
minimizing the quantity
\begin{equation}
  \chi_{N}^2 
  = 
  \frac{1}{M}\sum_{m=1}^{M}
  w_m\left(\rho_\tob(r_m) -
  \sum_{i=1}^{N} a_{N,i}\,\rho_{i}(r_m)\right)^2,
  \label{qpbest}
\end{equation}
which is a quadratic function of the coefficients $a_{N,i}$, consequently
defining a quadratic programming problem. The data points are given
equal weights by setting the constants $w_m=1/\rho_\tob^2(r_m)$.

To find such a set, we use an iterative algorithm
\citep{1989ApJ...343..113D}. With this method, a set of components is
successively built in $N$ steps. In the first step, $N_\tlib$
$\chi^2$-minimizations are performed using in turn each component of
the library. The base function that yields the lowest $\chi^2$-value
(hereafter denoted $\chi_{1}^2$) is retained as the first element of
our best-fitting set, with corresponding coefficient $a_{1,1}$. In
each next iteration, the functions in this set are preserved, while
the coefficients are allowed to vary. The set is subsequently extended
by adding the component from the library that yields the most
improvement of the fit, minimizing over all coefficients. In other
words, suppose we have obtained the best-fitting set of $N-1$ base
functions, with indices $n_1,\ldots,n_{N-1}$. Then, we add in turn the
remaining $N_\tlib-N+1$ components from the library, and calculate the
$N$ coefficients for each combination by means of the
$\chi^2$-minimizations
\begin{eqnarray}
  \chi_{(n_N)}^2 
  &=& 
  \min_{a_1,\ldots,a_N} 
  \frac{1}{M}\sum_{m=1}^{M}
  w_m\left(\rho_\tob(r_m) -
  \sum_{i=1}^{N} a_i\,\rho_{n_i}(r_m)\right)^2,\nonumber\\
  &&
  \mathrm{with}\ n_N \in \{1,\ldots,N_\tlib\} \backslash \{n_1,\ldots,n_{N-1}\}.
  \label{qpn}
\end{eqnarray}
From these $N_\tlib-N+1$ values, we determine the best fit
\begin{equation}
  \chi_N^2 = \chi_{(n_\tmin)}^2 = \min_{n_N} \chi_{(n_N)}^2,
  \label{qpnb}
\end{equation}
and add the base function with index $n_\tmin$ to the best-fitting
set, denoting $n_N=n_\tmin$. By renaming the indices of this set,
we thus obtain a linear combination of $N$ base functions
$\tilde\rho_i(\psi,r)$ with coefficients $a_{N,i}$ and a goodness of fit
$\chi_{N}^2$. The corresponding DF is then simply
\begin{equation}
  F(\calE,L) = \sum_{i=1}^{N} a_{N,i}\,F_i(\calE,L).
  \label{dfqp}
\end{equation}
The QP-algorithm allows additional linear constraints on the
coefficients. In particular, we impose upper and lower boundaries
\begin{equation}
  a_\mathrm{min} \leq a_{N,i} \leq a_\mathrm{max},
  \qquad \forall N\,;\ i=1,\ldots,N.
  \label{qp4}
\end{equation}
These constraints are not necessary, but they greatly reduce the
computational cost in the calculation of the DF. The reason for this
is straightforward: if the components are computed with numerical
errors $\delta_i F_i$ then the total numerical error of the DF is
\begin{equation}
  \delta F(\calE,L) 
  = 
  \frac{\sum_{i=1}^{N} |a|_{N,i}\,\delta_i F_i(\calE,L)}
  {\sum_{i=1}^{N} a_{N,i}}.
  \label{qpacc}
\end{equation}
The higher the absolute values of the coefficients $|a|_{N,i}$, the
smaller the errors $\delta_i F_i$ need to be to obtain a given $\delta
F$, which increases the computational time. Sensible boundary values
(\ref{qp4}) enable efficient calculations of the DF, while maintaining
satisfactory fits.

The resulting DF also needs to be physical, i.e.\ non-negative
everywhere in phase space. We found that all our DFs automatically
satisfy this condition without imposing explicit constraints.

This procedure has several advantages. The resulting DF remains
analytically tractable, which also simplifies the computation of all
the moments of this dynamical model. Furthermore, only a limited
number of data points are required, rather than the entire density
profile. The same algorithm can also be applied to data extracted from
simulations or observations, and a variety of different moments
besides the density can be used in the fitting procedure.

Finally, since all components have a priori the desired $\beta(r)$,
their linear combination will automatically generate the same
anisotropy profile. This is a very significant benefit. Indeed,
practice has shown that it is particularly difficult to construct
radially anisotropic systems from base functions that are too simple
(such as Fricke components, \citet{1952AN....280..193F}) with different
constant anisotropies. In that case an additional fitting is required
to the second-order moments. But more importantly, because radial
orbits influence the density both at small and large radii, the
summation of components with different anisotropies results into a
delicate fine-tuning to obtain adequate fits to both the density and
the velocity anisotropy. These problems are avoided if the components
already have the correct anisotropy profile, and although it is more
intricate to design such functions, this approach greatly reduces the
complexity of the quadratic programming procedure.

\subsection{The base functions}

We are now left with the construction of a library of adequate base
functions $\tilde\rho_i(\psi,r)$. The success of our modeling is
largely determined by this library.  As stated above, the components
have to generate the anisotropy profile (\ref{betagen}), while being
sufficiently simple to retrieve the corresponding DFs $F_i(\calE,L)$
from eq.~(\ref{rhodf}). In addition, the subsequent densities
$\rho_i(r_m)$ need to be able to reproduce the specific
characteristics of the given data to obtain a satisfactory fit.  In
the particular case of the Dehnen-McLaughlin halos (\ref{rhodehnen}),
this implies that the components should incorporate the central
density cusp and the asymptotic density slope at large radii, using
the given potential (\ref{psidehnen}).  In
\citetalias{2007A&A...471..419B} we derived a family of base functions
that meet all these requirements. Consider the family of augmented
densities
\begin{equation}
  \tilde\rho_i(\psi,r)
  =
  \rho_{0i}
  \left(\frac{\psi}{\psi_0}\right)^{p_i}
  \left(1-\frac{\psi^{s_i}}{\psi_0^{s_i}}\right)^{q_i}
  \left(\frac{r}{r_\txa}\right)^{-2\beta_0}
  \left(1+\frac{r^{2\delta}}{r_\txa^{2\delta}}\right)^{\beta_\delta},
\label{rhofam}
\end{equation}
where $\rho_{0i}$ are normalization constants such that
\begin{equation}
4\pi\int_0^{+\infty} \tilde\rho_i(\psi(r),r)r^2\,\txd r = 1,
\label{normalize}
\end{equation}
and $\psi_0$ denotes the depth of the (finite) potential well,
$\psi_0=\psi(0)$. The normalization constants are chosen such that the
sum of the coefficients of a good fit should approximate the total
mass of the input data, i.e.\ $\sum_{i=1}^{N} a_{N,i} \simeq
M_\mathrm{tot}$.

Apart from the four fixed parameters that determine the anisotropy
profile~(\ref{betadehnen}), the $f(\psi)$-part of these functions
contains three additional free parameters $p_i$, $q_i$ and $s_i$,
which respectively determine the asymptotic behavior at infinity, the
inner slope, and the transition rate between these two regions. They
satisfy the conditions $p_i + 2\beta_\infty > 3$, $q_i\leq 0$ and $s_i
>0$.

All other dynamical properties can be calculated from these augmented
densities. The augmented higher-order moments are derived in
Appendix~\ref{augmom.sec}, and we demonstrated in the Appendices of
\citetalias{2007A&A...471..419B} that for these $\rho_i(\psi,r)$ the
inverse Laplace-Mellin transforms can be performed analytically in
eq.~(\ref{rhodf}). The corresponding distribution functions can be
expressed as a series of Fox $H$-functions \citep{1961TAMS...98..395F}:
\begin{eqnarray}
  F_i(\calE,L)
  &=&
  \frac{\rho_{0i}}{M_{\mathrm{tot}}(2\pi\,\psi_0)^{3/2}}\,
  \nonumber \\
  &\times&
  \sum_{j=0}^\infty
  (-1)^j\,
  {q_i \choose j}\,
  \frac{\Gamma(1+p_i+js_i)}{\delta\,\Gamma(-{\beta_\delta})}\,
  \left(\frac{\calE}{\psi_0}\right)^{p_i+js_i-3/2}\,
  \nonumber \\
  &\times&
  H_{2,2}^{1,1}\!\!
  \left(\!\!
    \frac{L^2}{2r_\txa^2\calE}\!
    \left|\!
      \begin{Hmatrix}
        \left(1-\frac{\beta_\infty}{\delta},\frac{1}{\delta}\right)&
        \left(p_i+js_i-\frac{1}{2},1\right)
        \\[2mm]
        \left(-\frac{\beta_0}{\delta},\frac{1}{\delta}\right)&
        \left(0,1\right)
      \end{Hmatrix}
    \right.\!\!
  \right),
\label{dffoxH}
\end{eqnarray}
which can be written as a double series
\begin{eqnarray}
  F_i(\calE,L)
  &=&
  \frac{\rho_{0i}}{M_{\mathrm{tot}}(2\pi\,\psi_0)^{3/2}}\,
  \sum_{j=0}^\infty
  (-1)^j\,
  {q_i \choose j}\,
  \left(\frac{\calE}{\psi_0}\right)^{p_i+js_i-3/2}
  \nonumber \\
  &\times&
  \!\!\sum_{k=0}^\infty
  \!{\!\beta_\delta\! \choose \!k\!}\!
  \frac{\Gamma(1+p_i+js_i)}
  {\Gamma\left(p_i+js_i-\frac{1}{2}+\beta_k\right)\!
  \Gamma\left(\vhigh 1-\beta_k\right)}\!
  \left(\!\frac{L^2}{2r_\txa^2\calE}\!\right)^{-\beta_k},
\label{dfbase1}
\end{eqnarray}
where we used the auxiliary notation
\begin{equation}
  \beta_k
  =
  \left\{
    \begin{array}{ll}
      \beta_0-k\delta 
      \qquad &
      \mathrm{for}\quad L^2<2r_\txa^2\calE,\\
      \beta_\infty+k\delta 
      \qquad &
      \mathrm{for}\quad L^2>2r_\txa^2\calE.
    \end{array}
    \right.
\label{dfbase2}
\end{equation}
The double summation $\sum_j \sum_{k}$ can be computed by changing the
indices to $\sum_l \sum_{j+k=l}$, so that the inner summation becomes
a finite sum of $l+1$ terms for each value of $l$; the index $l$ is
increased until the total sum alters by less than a required numerical
error $\delta_i F_i$. Due to the double summation, the computational
time is an inverse quadratic function of $\delta_i F_i$.

We proved in \citetalias{2007A&A...471..419B} that these base DFs are
continuous and non-negative everywhere in physical phase
space. Moreover, we showed that they generate exactly the generalized
Plummer potential-density pairs \citep{1979AZh....56..976V}
\begin{eqnarray}
  \psi_\mathrm{gp}(r)
  &=&
  \frac{GM_\mathrm{tot}}{(r_\txs^\eta+r_{\phantom{\txs}}^\eta)^{1/\eta}},
  \label{veltpot}
  \\
  \rho_\mathrm{gp}(r)
  &=&
  \frac{(1+\eta)\,M_\mathrm{tot}}{4\pi}\,
  \frac{r_\txs^\eta}{
    r_{\phantom{\txs}}^{2-\eta}\,(r_\txs^\eta+r_{\phantom{\txs}}^\eta)^{2+1/\eta}},
  \label{veltrho}
\end{eqnarray}
with a linear $\beta-\gamma$ relation. Since these systems are closely
related to the Dehnen-McLaughlin halos, this is a promising result for
the success of the quadratic programming routine.

The {\sc Fortran} source code with the DF base functions and the
augmented moments is available on request.

\subsection{The library of components}

Every given Dehnen-McLaughlin halo requires a specific component
library. In particular, the parameters $p_i$, $q_i$ are constrained by
the potential. If we examine the asymptotic behavior of the
Dehnen-McLaughlin potential~(\ref{psidehnen}) in more detail, we find
\begin{equation}
\begin{array}{rclcl}
  \psi(r) &\sim& \displaystyle\psi_0-a\,r^{(11-10\beta_0)/9}+\cdots 
  &\mathrm{for}& r\rightarrow 0,
  \\
  \psi(r) &\sim& \displaystyle\frac{1}{r} 
  &\mathrm{for}& r\rightarrow \infty.
\end{array}
\end{equation}
Introducing these asymptotic expansions in the
expression~(\ref{rhofam}) we find for the inner and outer slopes of
the density
\begin{equation}
\begin{array}{rclcl}
  \tilde\rho_i(\psi(r),r) &\sim& \displaystyle
  r^{-2\beta_0+q_i(11-10\beta_0)/9}
  &\mathrm{for}& r\rightarrow 0,
  \\
  \tilde\rho_i(\psi(r),r) &\sim& \displaystyle
  r^{-2\beta_\infty-p_i}
  &\mathrm{for}& r\rightarrow \infty.
\end{array}
\end{equation}
Evidently, the parameters $p_i$ stipulate the density slope at large
radii. Because the models fall as $r^{-\gamma_\infty}$, no components
can be used in the fitting routine that fall less rapidly. Using
eq.~(\ref{gammadehneni}), this puts a boundary on the $p_i$,
\begin{equation}
  p_i
  \geq
  \frac{31-2\beta_0-18\beta_\infty}{9} \equiv p_\tmin(\beta_0,\beta_\infty).
  \label{pmin}
\end{equation}
Conversely, the density slope at small radii depends on the parameters
$q_i$. The density diverges toward the center as $r^{-\gamma_0}$, and
we cannot use components in the fitting routine that have a steeper
slope. Thus we obtain from eq.~(\ref{gammadehnen0})
\begin{equation}
  q_i
  \geq
  -\frac{7 - 8\beta_0}{11-10\beta_0} \equiv q_\tmin(\beta_0).
  \label{qmin}
\end{equation}
So if a fit to a halo has at least one component with parameter
$p_\tmin$ and one with $q_\tmin$, this fit has the same slope
as the given density at small and large radii.

Finally, the parameters $s_i$ have a similar role as $\delta$, in the
sense that they control the transition rate between the inner and
outer density slopes. Their value can be chosen freely, but we found
that excellent results are obtained with a single fixed value
\begin{equation}
s_i \equiv 2\delta = \eta,
\end{equation}
for all components. This is the same choice as the generalized Plummer
models in \citetalias{2007A&A...471..419B}. It also simplifies the
computation of the DFs~(\ref{dfbase1})-(\ref{dfbase2}), and it
further facilitates the fitting process, leaving only $p_i$ and $q_i$
as free parameters.

\begin{deluxetable*}{crrrrrrrrrr}
\tablecolumns{11}
\tablewidth{\textwidth}
\tablecaption{Components of the six QP-models.\label{models.tbl}}
\tablehead{
\colhead{} &
\colhead{1} & \colhead{2} & \colhead{3} & \colhead{4} & \colhead{5} & 
\colhead{6} & \colhead{7} & \colhead{8} & \colhead{9} & \colhead{10}\\[1mm] 
\multicolumn{11}{c}{$\vhigh\beta_0 = 0.0,\qquad \beta_\infty = 0.0,\qquad s= 4/9,\qquad \delta=2/9 
\qquad\qquad p_i = 3.4\overline{4}$, 4, 5, 7.5, 10
$\qquad\qquad q_i = -0.\overline{63}$, -0.5, -0.4, -0.3, -0.15, 0}}
\startdata

$a_{10,i}$  & $  0.2047^{\phn}$   & $ -0.0380^{\phn}$   & $ -0.5411^{\phn}$   & $  1.6652^{\phn}$   & $  0.0414^{\phn}$   & $ -3.8806^{\phn}$   & $  0.1312^{\phn}$   & $  2.5207^{\phn}$   & $ -0.2684^{\phn}$   & $  1.1652^{\phn}$   \\
$\rho_{0i}$ & $  0.0053^{\phn}$   & $  1.9783^{\phn}$   & $  0.0009^{\phn}$   & $  2.5540^{\phn}$   & $  0.4890^{\phn}$   & $  3.0677^{\phn}$   & $  0.0384^{\phn}$   & $  3.6728^{\phn}$   & $  6.1920^{\phn}$   & $  0.0008^{\phn}$   \\
$p_i$       & $  4.0000^{\phn}$   & $ 10.0000^{\phn}$   & $  3.4444^{\phn}$   & $ 10.0000^{\phn}$   & $  7.5000^{\phn}$   & $ 10.0000^{\phn}$   & $  5.0000^{\phn}$   & $ 10.0000^{\phn}$   & $ 10.0000^{\phn}$   & $  3.4444^{\phn}$   \\
$q_i$       & $ -0.6364^{\phn}$   & $ -0.6364^{\phn}$   & $  0.0000^{\phn}$   & $ -0.5000^{\phn}$   & $ -0.6364^{\phn}$   & $ -0.4000^{\phn}$   & $ -0.6364^{\phn}$   & $ -0.3000^{\phn}$   & $  0.0000^{\phn}$   & $ -0.6364^{\phn}$   \\[0.5mm]
$\chi^2_N$  & $ 0.44\cdot 10^{0 }$ & $ 0.17\cdot 10^{0 }$ & $ 0.67\cdot 10^{-1}$ & $ 0.33\cdot 10^{-1}$ & $ 0.24\cdot 10^{-2}$ & $ 0.17\cdot 10^{-2}$ & $ 0.53\cdot 10^{-4}$ & $ 0.36\cdot 10^{-4}$ & $ 0.48\cdot 10^{-5}$ & $ 0.82\cdot 10^{-7}$ \\

\cutinhead{$\vhigh\beta_0 = 0.0,\qquad \beta_\infty = 0.2,\qquad s= 4/9,\qquad \delta=2/9
\qquad\qquad p_i = 3.0\overline{4}$, 4, 5, 7.5, 10
$\qquad\qquad q_i = -0.\overline{63}$, -0.5, -0.4, -0.3, -0.15, 0}

$a_{10,i}$  & $ -0.1534^{\phn}$   & $ 15.5467^{\phn}$   & $ -0.0006^{\phn}$   & $  0.4549^{\phn}$   & $ -0.2707^{\phn}$   & $ 74.2329^{\phn}$   & $-88.6417^{\phn}$   & $  0.5880^{\phn}$   & $ -0.7724^{\phn}$   & $  0.0165^{\phn}$   \\
$\rho_{0i}$ & $  0.0238^{\phn}$   & $  0.0017^{\phn}$   & $  3.4087^{\phn}$   & $  4.4609^{\phn}$   & $ 11.3269^{\phn}$   & $  0.0014^{\phn}$   & $  0.0015^{\phn}$   & $  8.6356^{\phn}$   & $  5.4102^{\phn}$   & $  0.9797^{\phn}$   \\
$p_i$       & $  4.0000^{\phn}$   & $  3.0444^{\phn}$   & $ 10.0000^{\phn}$   & $ 10.0000^{\phn}$   & $ 10.0000^{\phn}$   & $  3.0444^{\phn}$   & $  3.0444^{\phn}$   & $ 10.0000^{\phn}$   & $ 10.0000^{\phn}$   & $  7.5000^{\phn}$   \\
$q_i$       & $ -0.6364^{\phn}$   & $  0.0000^{\phn}$   & $ -0.6364^{\phn}$   & $ -0.5000^{\phn}$   & $  0.0000^{\phn}$   & $ -0.6364^{\phn}$   & $ -0.5000^{\phn}$   & $ -0.1500^{\phn}$   & $ -0.4000^{\phn}$   & $ -0.6364^{\phn}$   \\[0.5mm]
$\chi^2_N$  & $ 0.48\cdot 10^{0 }$ & $ 0.27\cdot 10^{0 }$ & $ 0.43\cdot 10^{-1}$ & $ 0.93\cdot 10^{-2}$ & $ 0.13\cdot 10^{-2}$ & $ 0.88\cdot 10^{-3}$ & $ 0.10\cdot 10^{-4}$ & $ 0.95\cdot 10^{-5}$ & $ 0.48\cdot 10^{-5}$ & $ 0.32\cdot 10^{-7}$ \\

\cutinhead{$\vhigh\beta_0 = 0.0,\qquad \beta_\infty = 0.4,\qquad s= 4/9,\qquad \delta=2/9
\qquad\qquad p_i = 2.6\overline{4}$, 4, 5, 8, 12
$\qquad\qquad q_i = -0.\overline{63}$, -0.5, -0.4, -0.3, -0.15, 0}

$a_{10,i}$  & $  8.2375^{\phn}$   & $ -1.4318^{\phn}$   & $ -0.0467^{\phn}$   & $  0.2297^{\phn}$   & $  2.2705^{\phn}$   & $ -0.0111^{\phn}$   & $  0.1299^{\phn}$   & $ 10.3097^{\phn}$   & $ -0.3108^{\phn}$   & $-18.3771^{\phn}$   \\
$\rho_{0i}$ & $  0.0832^{\phn}$   & $  0.0032^{\phn}$   & $ 10.9019^{\phn}$   & $ 14.9246^{\phn}$   & $  0.0028^{\phn}$   & $  2.4301^{\phn}$   & $ 23.3335^{\phn}$   & $  0.1015^{\phn}$   & $ 18.6978^{\phn}$   & $  0.0934^{\phn}$   \\
$p_i$       & $  4.0000^{\phn}$   & $  2.6444^{\phn}$   & $ 12.0000^{\phn}$   & $ 12.0000^{\phn}$   & $  2.6444^{\phn}$   & $  8.0000^{\phn}$   & $ 12.0000^{\phn}$   & $  4.0000^{\phn}$   & $ 12.0000^{\phn}$   & $  4.0000^{\phn}$   \\
$q_i$       & $ -0.6364^{\phn}$   & $  0.0000^{\phn}$   & $ -0.6364^{\phn}$   & $ -0.5000^{\phn}$   & $ -0.6364^{\phn}$   & $ -0.6364^{\phn}$   & $ -0.3000^{\phn}$   & $ -0.4000^{\phn}$   & $ -0.4000^{\phn}$   & $ -0.5000^{\phn}$   \\[0.5mm]
$\chi^2_N$  & $ 0.48\cdot 10^{0 }$ & $ 0.18\cdot 10^{0 }$ & $ 0.28\cdot 10^{-1}$ & $ 0.93\cdot 10^{-2}$ & $ 0.42\cdot 10^{-2}$ & $ 0.14\cdot 10^{-3}$ & $ 0.36\cdot 10^{-4}$ & $ 0.13\cdot 10^{-4}$ & $ 0.68\cdot 10^{-5}$ & $ 0.24\cdot 10^{-6}$ \\

\cutinhead{$\vhigh\beta_0 = 0.0,\qquad \beta_\infty = 0.6,\qquad s= 4/9,\qquad \delta=2/9
\qquad\qquad p_i = 2.2\overline{4}$, 3.5, 5, 8, 12
$\qquad\qquad q_i = -0.\overline{63}$, -0.5, -0.4, -0.3, -0.15, 0}

$a_{10,i}$  & $  0.3552^{\phn}$   & $  2.0738^{\phn}$   & $ -0.0094^{\phn}$   & $  0.0086^{\phn}$   & $ 16.3498^{\phn}$   & $-17.4861^{\phn}$   & $  0.0731^{\phn}$   & $ -0.3019^{\phn}$   & $ -0.2109^{\phn}$   & $  0.1477^{\phn}$   \\
$\rho_{0i}$ & $  0.1222^{\phn}$   & $  0.0063^{\phn}$   & $ 16.7922^{\phn}$   & $ 23.2672^{\phn}$   & $  0.0053^{\phn}$   & $  0.0055^{\phn}$   & $  4.3156^{\phn}$   & $  0.1372^{\phn}$   & $  8.0592^{\phn}$   & $ 10.5109^{\phn}$   \\
$p_i$       & $  3.5000^{\phn}$   & $  2.2444^{\phn}$   & $ 12.0000^{\phn}$   & $ 12.0000^{\phn}$   & $  2.2444^{\phn}$   & $  2.2444^{\phn}$   & $  8.0000^{\phn}$   & $  3.5000^{\phn}$   & $  8.0000^{\phn}$   & $  8.0000^{\phn}$   \\
$q_i$       & $ -0.6364^{\phn}$   & $  0.0000^{\phn}$   & $ -0.6364^{\phn}$   & $ -0.5000^{\phn}$   & $ -0.6364^{\phn}$   & $ -0.5000^{\phn}$   & $ -0.6364^{\phn}$   & $ -0.5000^{\phn}$   & $ -0.3000^{\phn}$   & $ -0.1500^{\phn}$   \\[0.5mm]
$\chi^2_N$  & $ 0.44\cdot 10^{0 }$ & $ 0.15\cdot 10^{0 }$ & $ 0.18\cdot 10^{-1}$ & $ 0.67\cdot 10^{-2}$ & $ 0.31\cdot 10^{-2}$ & $ 0.16\cdot 10^{-4}$ & $ 0.13\cdot 10^{-4}$ & $ 0.10\cdot 10^{-4}$ & $ 0.17\cdot 10^{-5}$ & $ 0.20\cdot 10^{-6}$ \\

\cutinhead{$\vhigh\beta_0 = 0.0,\qquad \beta_\infty = 0.8,\qquad s= 4/9,\qquad \delta=2/9
\qquad\qquad p_i = 1.8\overline{4}$, 3, 5, 8, 12
$\qquad\qquad q_i = -0.\overline{63}$, -0.5, -0.4, -0.3, -0.15, 0}

$a_{10,i}$  & $ -0.0321^{\phn}$   & $ -6.1421^{\phn}$   & $  0.0008^{\phn}$   & $ -0.0013^{\phn}$   & $ 24.5830^{\phn}$   & $-100.0000^{\phn}$  & $ -0.0163^{\phn}$   & $ -0.0926^{\phn}$   & $ 40.5233^{\phn}$   & $ 42.1774^{\phn}$   \\
$\rho_{0i}$ & $  0.1791^{\phn}$   & $  0.0121^{\phn}$   & $ 25.3820^{\phn}$   & $ 35.5933^{\phn}$   & $  0.0101^{\phn}$   & $  0.0109^{\phn}$   & $  2.3877^{\phn}$   & $  0.2367^{\phn}$   & $  0.0105^{\phn}$   & $  0.0116^{\phn}$   \\
$p_i$       & $  3.0000^{\phn}$   & $  1.8444^{\phn}$   & $ 12.0000^{\phn}$   & $ 12.0000^{\phn}$   & $  1.8444^{\phn}$   & $  1.8444^{\phn}$   & $  5.0000^{\phn}$   & $  3.0000^{\phn}$   & $  1.8444^{\phn}$   & $  1.8444^{\phn}$   \\
$q_i$       & $ -0.6364^{\phn}$   & $  0.0000^{\phn}$   & $ -0.6364^{\phn}$   & $ -0.5000^{\phn}$   & $ -0.6364^{\phn}$   & $ -0.4000^{\phn}$   & $ -0.3000^{\phn}$   & $ -0.3000^{\phn}$   & $ -0.5000^{\phn}$   & $ -0.1500^{\phn}$   \\[0.5mm]
$\chi^2_N$  & $ 0.40\cdot 10^{0 }$ & $ 0.11\cdot 10^{0 }$ & $ 0.11\cdot 10^{-1}$ & $ 0.46\cdot 10^{-2}$ & $ 0.22\cdot 10^{-2}$ & $ 0.66\cdot 10^{-5}$ & $ 0.26\cdot 10^{-5}$ & $ 0.13\cdot 10^{-5}$ & $ 0.47\cdot 10^{-6}$ & $ 0.24\cdot 10^{-6}$ \\

\cutinhead{$\vhigh\beta_0 = 0.0,\qquad \beta_\infty = 1.0,\qquad s= 4/9,\qquad \delta=2/9
\qquad\qquad p_i = 1.4\overline{4}$, 3, 5, 8, 12
$\qquad\qquad q_i = -0.\overline{63}$, -0.5, -0.4, -0.3, -0.15, 0}

$a_{10,i}$  & $ -5.2047^{\phn}$   & $  0.1086^{\phn}$   & $ 62.6962^{\phn}$   & $ 40.1800^{\phn}$   & $ -0.0002^{\phn}$   & $  0.0003^{\phn}$   & $-96.6969^{\phn}$   & $ -0.1234^{\phn}$   & $  0.0273^{\phn}$   & $  0.0128^{\phn}$   \\
$\rho_{0i}$ & $  0.0192^{\phn}$   & $  2.9967^{\phn}$   & $  0.0201^{\phn}$   & $  0.0213^{\phn}$   & $ 37.6938^{\phn}$   & $ 44.2497^{\phn}$   & $  0.0207^{\phn}$   & $  3.7237^{\phn}$   & $  7.8118^{\phn}$   & $  0.8543^{\phn}$   \\
$p_i$       & $  1.4444^{\phn}$   & $  5.0000^{\phn}$   & $  1.4444^{\phn}$   & $  1.4444^{\phn}$   & $ 12.0000^{\phn}$   & $  8.0000^{\phn}$   & $  1.4444^{\phn}$   & $  5.0000^{\phn}$   & $  5.0000^{\phn}$   & $  3.0000^{\phn}$   \\
$q_i$       & $ -0.6364^{\phn}$   & $ -0.6364^{\phn}$   & $ -0.5000^{\phn}$   & $ -0.3000^{\phn}$   & $ -0.6364^{\phn}$   & $  0.0000^{\phn}$   & $ -0.4000^{\phn}$   & $ -0.5000^{\phn}$   & $  0.0000^{\phn}$   & $ -0.1500^{\phn}$   \\[0.5mm]
$\chi^2_N$  & $ 0.35\cdot 10^{0 }$ & $ 0.36\cdot 10^{-1}$ & $ 0.10\cdot 10^{-1}$ & $ 0.24\cdot 10^{-3}$ & $ 0.41\cdot 10^{-4}$ & $ 0.16\cdot 10^{-4}$ & $ 0.13\cdot 10^{-4}$ & $ 0.45\cdot 10^{-6}$ & $ 0.89\cdot 10^{-7}$ & $ 0.22\cdot 10^{-8}$ \\
\enddata

\tablecomments{Our six models are determined by the anisotropy
parameters $\beta_0=0$ and $\beta_\infty$. With the QP-algorithm, we
built linear combinations up to 10 components, each characterized by
$s$, $\delta$, $p_i$, $q_i$ and $\rho_{0i}$. The parameters $s$ and
$\delta$ are the same for each component, while $p_i$ and $q_i$ are
selected from a library of 30 components. The five $p_i$ values and
six $q_i$ values in each library are given in the headers.  Every
$N\,$th column lists the value $\chi^2_N$ of the best fit with the
first $N$ components. The coefficients are given for the final fit
with 10 components, i.e.\ $a_{10,i}$.}

\end{deluxetable*}

\begin{figure}
\centering
\plotone{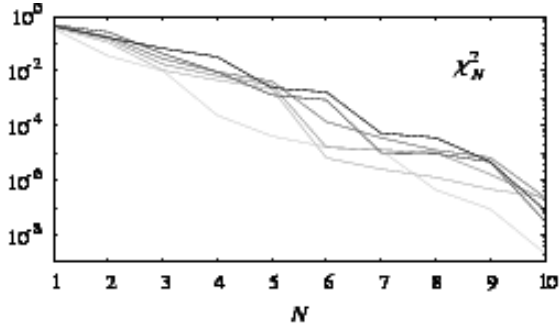}
\caption{The obtained $\chi_N^2$ for the six QP-models, explicitly as
a function of the number of components in the fit. The different
curves correspond to $\beta_\infty=0$, 0.2, 0.4, 0.6, 0.8 and 1, with
a grayscale ranging from black ($\beta_\infty=0$) to lightgray
($\beta_\infty=1$).}
\label{f1.eps}
\end{figure}

\section{Results}
\label{results.sec}

\subsection{The minimization}

Now that we have derived the necessary mathematical tools, we can
present the results for the Dehnen-McLaughlin halos. Without loss of
generality, we can work in dimensionless units $G =
M_{\mathrm{tot}}=r_\txs=r_\txa=1$, and we limit ourselves to
$\varepsilon=3$. Consequently, the models are determined by the
anisotropy parameters $\beta_0$ and $\beta_\infty$. Although we are
able to generate models with arbitrary values for these parameters,
realistic dark matter halos are nearly isotopic near the center and
radially anisotropic at large radii, so that we concentrate on six
representative models with $\beta_0=0$ and $\beta_\infty=0$, 0.2, 0.4,
0.6, 0.8, 1. We verified that the modeling procedure works equally
well for models with non-zero values of $\beta_0$. Finally, it is
evident from eq.~(\ref{etadehnen}) that $\beta_0=0$ sets the
parameters $s_i \equiv 2\delta = \eta = 4/9$.

As we demonstrated above, the very specific form of the base functions
(\ref{rhofam}) simplifies our QP-algorithm considerably for these
models. Only the parameters $p_i$ and $q_i$ remain to construct a
library of components, and we have found that only 30 components are
sufficient to yield excellent fits. The parameters $p_i$ take five
values, ranging from $p_\tmin(0,\beta_\infty)$ to 10 or 12, depending
on the model. The parameters $q_i$ take six values from $q_\tmin(0)$
to $0$. We list these values for each model in the headers of
Table~\ref{models.tbl}.

We extract $M=25$ values of the density~(\ref{rhodehnen}) at radii
$r_m$, distributed logarithmically between \mbox{$10^{-3}\,r_\txs$}
and \mbox{$10^4\,r_\txs$}, that serve as input data $\rho_\tob(r_m)$
in each QP-procedure. Evidently, this range is much larger than the
virialized region in $N$-body simulations. This larger range is
therefore not intended to be realistic, but rather to demonstrate that
our models are accurate to arbitrary distances. Furthermore, this
makes it possible to create discrete equilibrium systems from the DFs,
by means of Monte Carlo simulators, that trace very closely the
Dehnen-McLaughlin halos. After calculating the
densities~(\ref{rhofam}) of every library component at these radii
$\tilde\rho_i(\psi(r_m),r_m)$, we can perform the QP-algorithm for the
six values of $\beta_\infty$, constructing iteratively the
best-fitting linear combination~(\ref{qpbest}) of $N$ components with
additional constraints of the form (\ref{qp4}),
\begin{equation}
-100 \leq a_{N,i} \leq 100, \qquad \forall N\,;\ i=1,\ldots,N.
\label{qp4b}
\end{equation}

We show the results for the six models in Table~\ref{models.tbl}. The
columns list the components of the fits.  Every $\chi_N^2$ denotes the
goodness of fit of the best linear combination of the components in
columns $1$ to $N$. Each component is determined by the parameters
$p_i$, and $q_i$, which in turn define the normalization constants
$\rho_{0i}$ from eq.~(\ref{normalize}). The coefficients are only
given for the final fit with 10 components, i.e.\ $a_{10,i}$.  It can
be checked that for each model
\begin{equation}
\sum_{i=1}^{10} a_{10,i} \simeq M_\mathrm{tot} = 1,
\end{equation}
indicating excellent fits. Combining this result with
eqs.~(\ref{qpacc}) and (\ref{qp4b}), it can be seen that if $N=10$, the
numerical errors of the base functions $\delta_i F_i$ need at most be
a factor $10^3$ smaller than a given error $\delta F$, allowing
efficient computations of the DF with sufficient accuracy.

\begin{figure}
\centering
\plotone{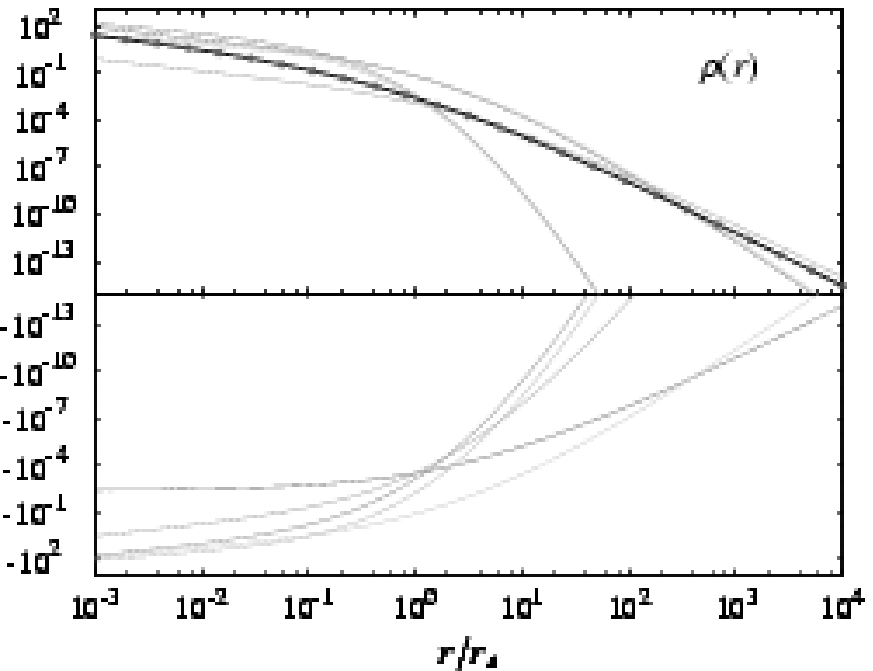}
\caption{The 10 individual components of the fitted density for the
QP-model with $\beta_0=0$ and $\beta_\infty=0.4$. Their sum is the
QP-density (black thick curve), fitting the 25 data points (dots).}
\label{f2.eps}
\end{figure}

The resulting $\chi_N^2$ for each model are also displayed in
Fig.~\ref{f1.eps}. Evidently, $N=10$ components are more than
sufficient to obtain very accurate dynamical models. As an example,
Fig.~\ref{f2.eps} shows the 10 individual components of the QP-model
with $\beta_\infty=0.4$.  Although this fit has the highest
$\chi_{10}^2$ of our set, its total density is a very close
approximation to the given data over the entire range in radius.

\subsection{The moments}

\begin{figure*}
\centering
\plotone{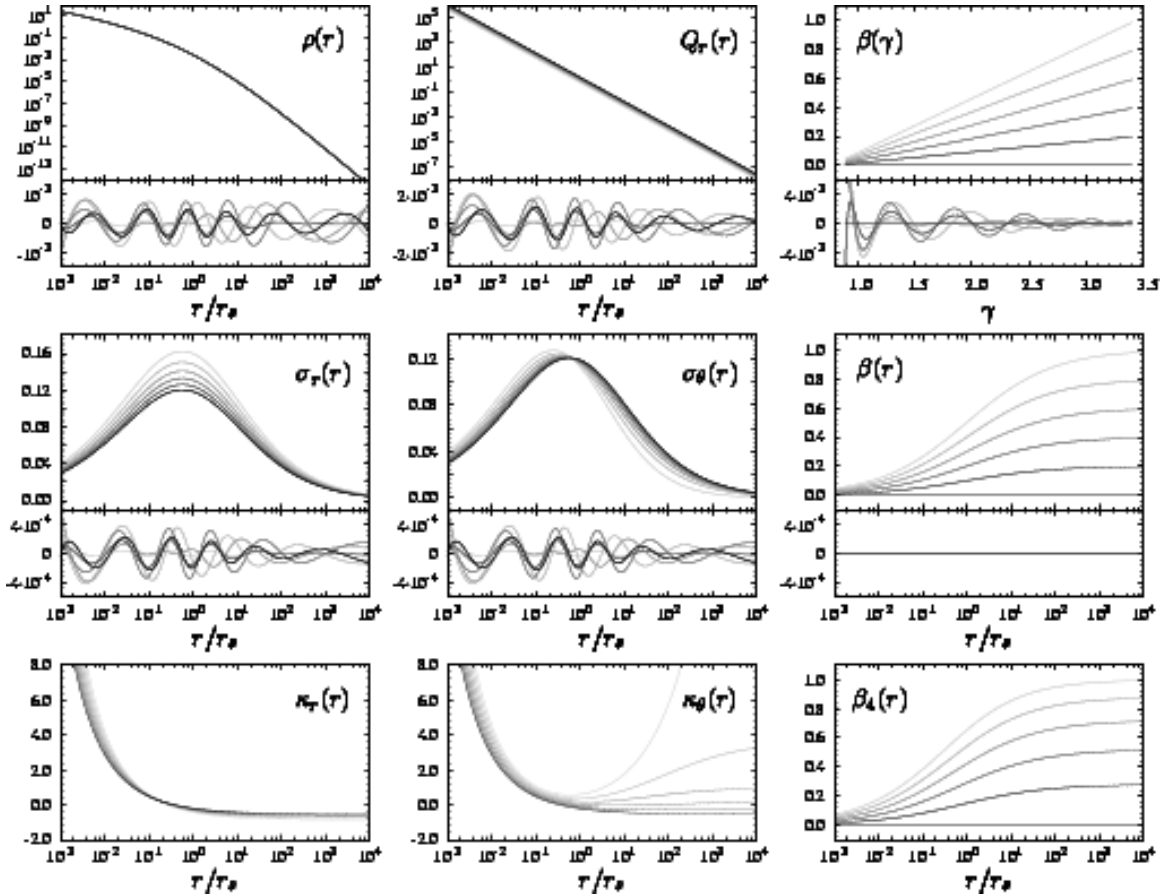}
\caption{The most important moments for our set of representative
  models with 10 components. Top row: the density $\rho(r)$, the
  pseudo phase-space density $\pseudo_r(r)$ and the $\beta-\gamma$
  relation. Below each graph, the relative errors with respect to the
  theoretical profiles is shown.  Middle row: the radial velocity
  dispersion $\sigma_r(r)$, the tangential velocity dispersion
  $\sigma_\theta(r)$, and the anisotropy $\beta(r)$, also with the
  relative errors. Bottom row: the radial kurtosis $\kappa_r(r)$, the
  tangential kurtosis $\kappa_\theta(r)$, and fourth-order anisotropy
  $\beta_4(r)$. The models and grayscaling are the same as in
  Fig.~\ref{f1.eps}.}
\label{f3.eps}
\end{figure*}

Fig.~\ref{f3.eps} displays several moments for our six models,
with 10 components.  The top row shows the density $\rho(r)$, the
pseudo phase-space density $\pseudo_r(r)$, and the $\beta-\gamma$
relation. Below each graph, we calculated the residual errors between
the QP-fits and the theoretical curves, i.e.\ for each profile $f(r)$
we have
\begin{equation}
  \Delta f(r) 
  = 
  \frac{f_\tob(r)-f_\tqp(r)}{f_\tob(r)}.
\end{equation}
As can be seen, the relative errors on the densities are less than
$10^{-3}$ along 7 orders of magnitude in radius, and the exact
asymptotic slopes of the models ensure excellent fits even beyond this
range. The power-law trend of $\pseudo_r(r)$ and the $\beta-\gamma$
relations are also reproduced very accurately with errors $\sim
10^{-3}$. Note that the small offset between the pseudo phase-space
density profiles for the different models is due to the dependence of
$\sigma_r(r)$ on $\beta_\infty$.

In the central row, we display the velocity dispersion profiles
$\sigma_r(r)$, $\sigma_\theta(r)$, and the anisotropies $\beta(r)$. It
is striking that, while the dispersion data were not used in the fit, the
deviations of these moments from the theoretical values are even
smaller, less than $5\cdot 10^{-4}$. Evidently, since the models have the
anisotropies~(\ref{betadehnen}) by construction, the $\beta(r)$
profiles are exact, without errors. Note also that all tangential
velocity dispersion profiles $\sigma_\theta(r)$ intersect at a common
radius $r=r_\txs \left(9/11\right)^{1/\eta}$.

While the density and dispersions are defined by the Dehnen-McLaughlin
halos, the higher-order moments are determined by the QP-models. The
fourth-order moments (see Appendix~\ref{augmom.sec}) allow us to
derive the radial and tangential kurtosis,
\begin{eqnarray}
  \kappa_r(r) 
  &=&
  \frac{\langle v_r^4 \rangle}{\sigma^4_r}(r) - 3,
  \\
  \kappa_\theta(r) 
  &=& 
  \frac{\langle v_\theta^4 \rangle}{\sigma^4_\theta}(r) - 3,
\end{eqnarray}
and the fourth-order anisotropy
\begin{equation}
  \beta_4(r)
  =
  1 - \frac{\langle v_\theta^4 \rangle}{\langle v_r^4 \rangle}(r).
\end{equation}
Interestingly, as a result of the separable form of the augmented
densities, we find that the $\beta_4(r)$ profiles are only a function
of the $\beta(r)$,
\begin{equation}
  \beta_4(r)
  =
  \frac{1}{2}\beta(r)\left(3-\beta(r)\vhighs\right) +
  \frac{1}{2\beta_\delta} \left(\beta_0-\beta(r)\vhighs\right)
  \left(\beta_\infty-\beta(r)\vhighs\right).
\end{equation}
These profiles are shown in the bottom row of Fig.~\ref{f3.eps}.  The
kurtosis values describe the non-Gaussianity of the velocity
distributions at a certain radius. Our radial kurtosis values are
very large in the center, which indicates that the $v_r$-distributions
are very peaked (leptokurtic) at small radii. The $\kappa_r(r)$ curves
decrease rapidly as a function of radius: they reach zero at radii
between 0.26-0.36 and become negative at larger radii, leading to
flat-topped (platykurtic) radial velocity distributions.  This
behavior is in accordance with $N$-body simulations
\citep{2004ApJ...601...37K,2005MNRAS.361L...1W}.  Clearly, the
value of $\beta_\infty$ has little influence on the radial kurtosis,
as in the case of $\pseudo_r(r)$. In contrast, the tangential kurtosis
$\kappa_\theta(r)$ curves do depend significantly on
$\beta_\infty$. All $v_\theta$-distributions are highly peaked at
small radii. For $\beta_\infty<0.4$ the tangential kurtosis decreases
to slightly negative values, i.e.\ at larger radii the tangential
velocity distributions become slightly flat-topped. Models with
$\beta_\infty\approx0.4$ have nearly Gaussian $v_\theta$-distributions
for $r>r_\txs$. If $\beta_\infty>0.4$, the $\kappa_\theta(r)$ profiles
reach a minimum value and increase again for larger radii.

\subsection{The distribution functions}

\begin{figure*}
\centering
\plotone{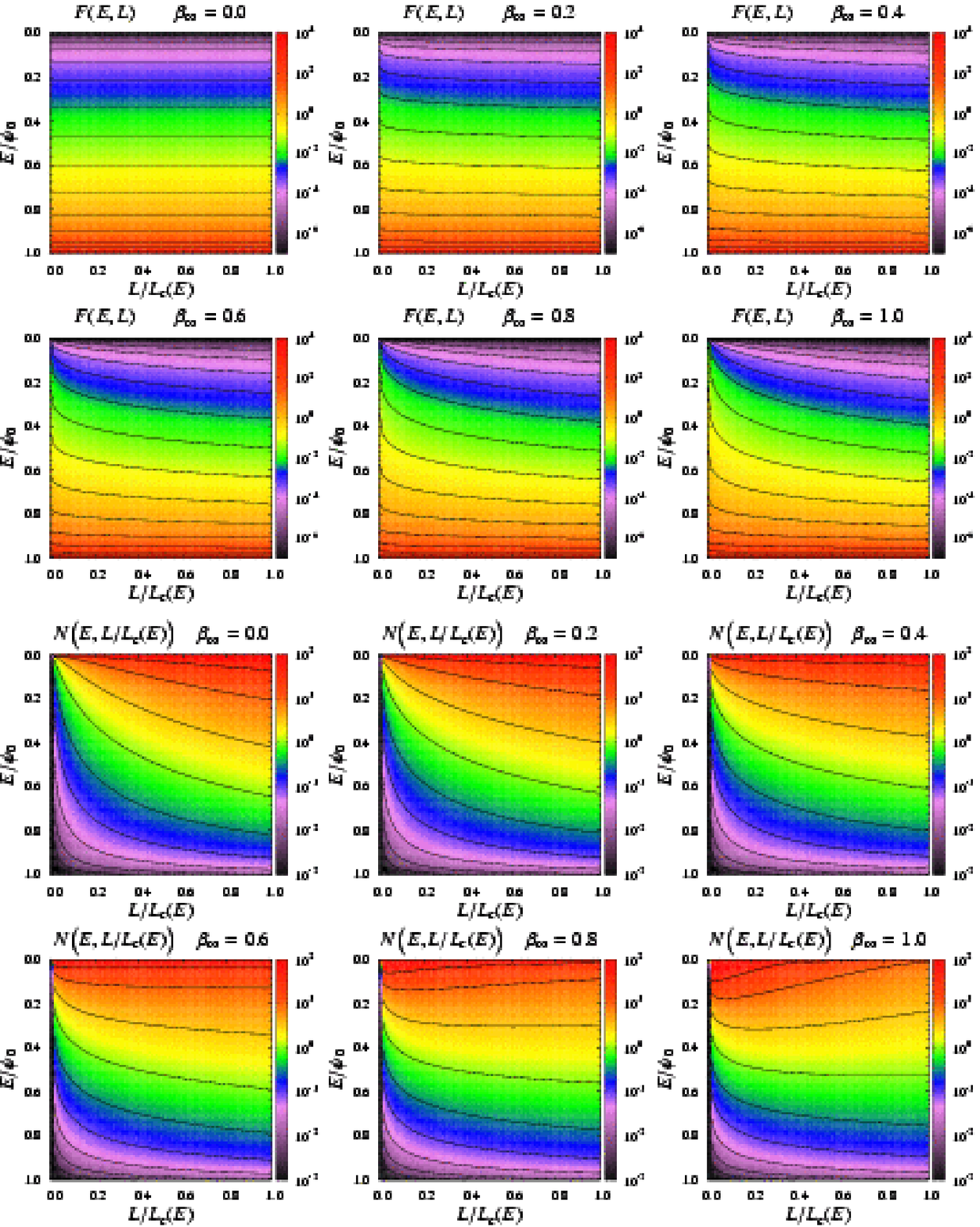}
\caption{The phase-space distribution functions for our set of models
  with 10 components. The six top panels show $F(\calE,L)$ as
  isoprobability contours in the integral space. The energy is scaled
  to the central potential and the angular momentum is scaled to the
  angular momentum $L_\txc(\calE)$ of a circular orbit with energy
  $\calE$. The contour levels and the coloring are scaled
  logarithmically.  The six bottom panels show the orbital
  distribution functions $N\left(\calE,L/L_\txc(\calE)\right)$
  for the same models.}
\label{f4.eps}
\end{figure*}

The six top panels of Fig.~\ref{f4.eps} show the distribution
functions $F(\calE,L)$ of our radially anisotropic systems with
$\beta_0=0$ and $\beta_\infty=0,\ldots,1$, expressed as logarithmic
isoprobability contours and a logarithmic color gradient in the
integral space, with $L$ scaled to $L_\txc(\calE)$, denoting the
angular momentum of a circular orbit with energy $\calE$. All models
are clearly physical, i.e.\ the DFs are non-negative everywhere. This
means that the Dehnen-McLaughlin Jeans models can actually be realized
by full dynamical models. Moreover, contrary to the Osipkov-Merritt
models, these functions fill the entire integral space.  In the
isotropic case, the contours are horizontal (no dependence on angular
momentum), and their orientation alters gradually with increasing
$\beta_\infty$ in an intuitive way, as orbits with high eccentricities
(i.e.\ low angular momentum) become more abundant.

The DFs describe the probability distributions of particles in phase
space, but not in the integral space. It is more physically meaningful
to consider the true orbital distributions
\begin{equation}
  N\left(\calE,L/L_\txc(\calE)\right)
  =
  F(\calE,L)\,g(\calE,L)\,L_\txc(\calE),
\end{equation}
with the so-called ''density of states'' function
\begin{equation}
  g(\calE,L) 
  = 
  16\pi^2 L \int_{\rp}^{\ra}\frac{\txd r}{\sqrt{2(\psi(r)-\calE)-L^2/r^2}},
\end{equation}
where $\rp$ and $\ra$ are the pericenter and apocenter of an orbit
with energy $\calE$ and angular momentum $L$. In other words, the
functions $N\left(\calE,L/L_\txc(\calE)\right)$ express the likelihood
of an orbit with energy $\calE$ and scaled angular momentum
$L/L_\txc(\calE)$.

The results are displayed in the bottom panels of Fig.~\ref{f4.eps} as
logarithmic isoprobability contours and a logarithmic color gradient
in the integral space. For high binding energies, all orbital
distributions contain increasing probabilities toward circular orbits
(high angular momentum). For low binding energies, i.e.\ at large
radii, the abundance of orbits with low angular momentum gradually
increases from the isotropic case to models with high values of
$\beta_\infty$.

\begin{figure*}
\centering
\plotone{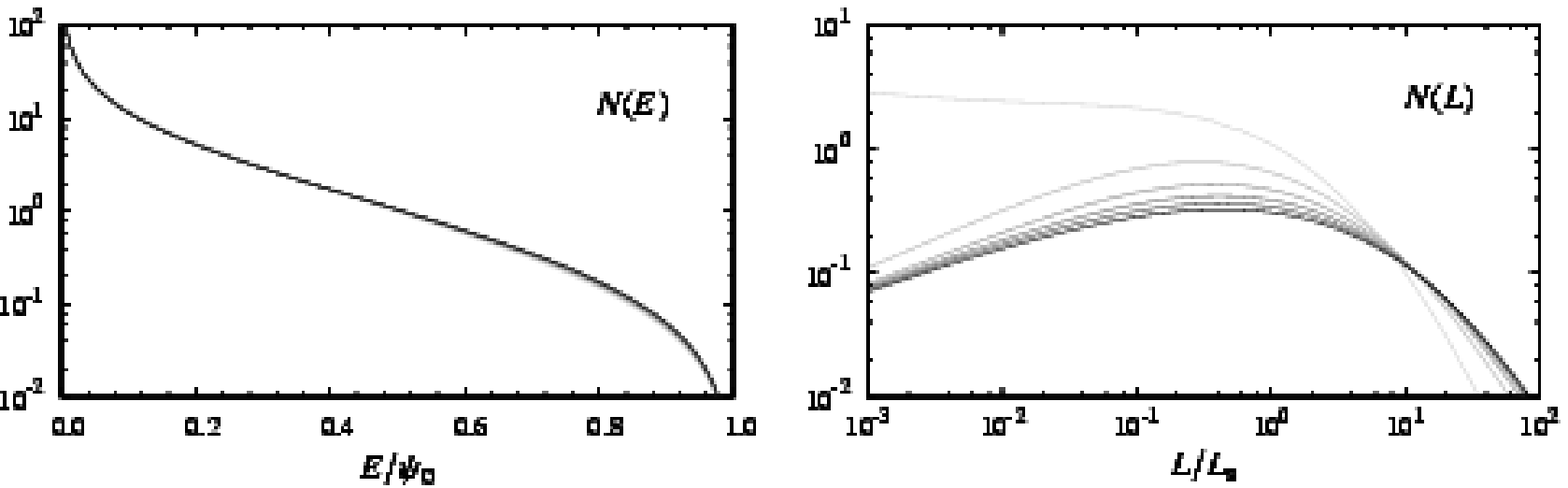}
\caption{The energy and angular momentum distributions of the six
QP-models with 10 components. The angular momenta are scaled to the
values $L_\txs$ of a circular orbit with radius $r_\txs$. The models
and grayscaling are the same as in Fig.~\ref{f1.eps}.}
\label{f5.eps}
\vspace{1cm}
\end{figure*}

\subsection{The marginal distributions}

We conclude the discussion of our Dehnen-McLaughlin DFs with an
analysis of the marginal distributions. The differential energy and
angular momentum distributions are the integrals of the orbital
distributions,
\begin{eqnarray}
N(\calE) &=& \int_0^{L_\txc(\calE)} N(\calE,L)\,\txd L,\\
N(L)     &=& \int_0^{\calE_\txc(L)} N(\calE,L)\,\txd \calE,
\end{eqnarray}
with
\begin{equation}
  N(\calE,L) 
  = 
  \frac{1}{L_\txc(\calE)}N\left(\calE,L/L_\txc(\calE)\right).
\end{equation}
These curves are displayed in Fig.~\ref{f5.eps}. The differential
energy distributions are all monotonously decreasing functions of
$\calE$. It is striking that these profiles are almost identical,
regardless of the anisotropy $\beta_\infty$. This result reinforces
previous dynamical studies \citep{1982MNRAS.200..951B}, and suggests
that $\rho(r)$, $\pseudo_r(r)$ and $\kappa_r(r)$ are linked to a
universal differential energy distribution, independent of
$\beta_\infty$, caused by the same physical processes.

In contrast, the angular momentum distributions depend on
$\beta_\infty$, most notably for high radial anisotropies.  For
increasing values of $\beta_\infty$, the fraction of orbits with low
angular momentum increases. \citet{2001ApJ...555..240B} proposed a
universal form for the integrated angular momentum distribution in
dark matter halos $M(L)$. Alternatively, \citet{2005ApJ...628...21S}
found a differential distribution
\begin{equation}
N_\mathrm{ss}(L) = \frac{1}{L_\txd^a \Gamma(a)}L^{a-1}\txe^{-L/L_\txd}. 
\label{angularss}
\end{equation}
Our models indicate a similar profile, with $a>0.9$, although the
functions~(\ref{angularss}) fall steeper than ours as $L$ increases.

\section{Conclusions}
\label{conclusions.sec}

In this paper we presented a set of anisotropic distribution functions
$F(\calE,L)$ for the Dehnen-McLaughlin Jeans models. We constructed
these DFs as a linear combination of base functions, which we fitted
to the halo density by means of a quadratic programming algorithm.
The base functions were specifically designed for this task, derived
from a separable augmented density that generates exactly the general
four-parameter velocity anisotropy profiles~(\ref{betagen}). We
demonstrated that the resulting fits are very accurate, from the
center to large radii, far beyond the range of $N$-body simulations.

This method has several advantages. The DFs can be written as a sum of
a double series, without numerical integrations or
inversions. Consequently, the DFs and all subsequent moments can be
computed with high accuracy. Moreover, the advanced form of the base
functions makes the QP-fitting very fast, requiring only a small
number of library components.

In this manner, we have constructed a family of dynamical models that
incorporate the observed features in $N$-body simulations. We
summarize their properties:
\begin{enumerate}
\item{The models a priori have the universal properties encountered in
    $N$-body studies of dark matter halos, which formed the building
    bricks of the Jeans models by \citet{2005MNRAS.363.1057D}: they
    generate a universal density profile, a power-law pseudo
    phase-space density $\pseudo_r(r)$, four-parameter velocity
    anisotropy profiles with arbitrary values of $\beta_0$ and
    $\beta_\infty$, and a linear $\beta-\gamma$ relation. In
    particular, we analyzed six models that are isotropic in the
    center and radially anisotropic at large radii.}
\item{The DFs are physical, i.e.\ they are non-negative over the
    entire phase space. This means that the Dehnen-McLaughlin Jeans
    models can actually be realized by self-consistent dynamical
    systems. In addition, the DFs are continuous and smooth functions
    and, contrary to the popular Osipkov-Merritt models, they fill the
    entire accessible phase space.}
\item{The energy distributions are monotonously decreasing functions,
    and like the radial kurtosis profiles, these functions are nearly
    independent of $\beta_\infty$. This suggests that $\rho(r)$,
    $\pseudo_r(r)$, $\kappa_r(r)$, and $N(\calE)$ have a universal
    form, caused by the same physical processes.}
\end{enumerate}
In addition, we have also been able to generate dynamical models with
a non-linear $\beta-\gamma$ relation, i.e.\ $\delta \neq \eta/2$ and
$r_\txa \neq r_\txs$, albeit with higher $\chi_N^2$ values.  From
these successful results for the Dehnen-McLaughlin halos, we expect
equally adequate fits for other Zhao models, such as the NFW
halos. Other profiles, such as the S\'ersic-type densities, might
require a modified family of base functions.  Such a systematic study
could unravel more hitherto hidden properties of dark matter halos,
and the various connections between these characteristics.

Our set of dynamical models is not only useful for a purely
theoretical analysis. The DFs can also serve to generate initial
conditions with Monte Carlo simulators
\citep[e.g.][]{2004ApJ...601...37K,2007MNRAS.375.1157B} to investigate
the physical processes within these equilibrium models by means of
controlled numerical $N$-body simulations. Finally, our algorithm and
base functions can be used in other dynamical studies, using different
data moments than the density. This could lead to a significant
improvement in the dynamical modeling of observed stellar systems,
such as clusters of galaxies.

\appendix
\section{Derivation of the augmented moments}
\label{augmom.sec}

In a spherical dynamical model, the anisotropic velocity moments of
the DF are
\begin{equation}
  \mu_{2n,2m}(r)
  =
  2\pi M_\mathrm{tot} \int_{-\infty}^{+\infty}\txd v_r \int_0^{+\infty}
  F(\calE,L)\,v_r^{2n}\,v_T^{2m+1}\,\,\txd v_T.
\end{equation}
Using the augmented density formalism, these moments can be calculated
as $\mu_{2n,2m}(r) = \tilde{\mu}_{2n,2m}(\psi(r),r)$, with
\begin{equation}
    \tilde{\mu}_{2n,2m}(\psi,r)
    =
    \frac{2^{m+n}}{\sqrt{\pi}}\,
    \frac{\Gamma\left(n+\frac{1}{2}\right)}
    {\Gamma\left(m+n\right)}
    \int_0^\psi
    \left(\psi-\psi'\right)^{m+n-1}
    D_{r^2}^m\left[r^{2m}\tilde\rho(\psi',r)\right]
    \txd\psi',
\end{equation}
where $D_x^m$ denotes the $m\,$th differentiation with respect to
$x$. Applying this to our base functions~(\ref{rhofam}), we obtain
for the augmented second-order moments
\begin{equation}
  \tilde\mu_{20,i}(\psi,r)
  =
  \frac{\rho_{0i}\psi_0}{s_i}
  \left(\frac{r}{r_\txa}\right)^{-2\beta_0}
  \left(1+\frac{r^{2\delta}}{r_\txa^{2\delta}}\right)^{\beta_\delta} 
  B_{y}\left(\frac{1+p_i}{s_i},1+q_i\right),
\label{mu20fam}
\end{equation}
with $y = (\psi/\psi_0)^{s_i}$, and
\begin{equation}
  \tilde\mu_{02,i}(\psi,r) 
  =
  2\left(1-\beta(r)\vhighs\right)\,\tilde\mu_{20,i}(\psi,r).
\end{equation}
The augmented fourth-order moments have the form
\begin{eqnarray}
  \tilde\mu_{40,i}(\psi,r)
  &=&
  \frac{3\rho_{0i}\psi_0^2}{s_i}
  \left(\frac{r}{r_\txa}\right)^{-2\beta_0}
  \left(1+\frac{r^{2\delta}}{r_\txa^{2\delta}}\right)^{\beta_\delta} 
  \left[
  \frac{\psi}{\psi_0}
  B_{y}\left(\frac{1+p_i}{s_i},1+q_i\right)
  -
  B_{y}\left(\frac{2+p_i}{s_i},1+q_i\right)
  \right],
  \label{mu40fam}\nonumber\\
  \\
  \tilde\mu_{22,i}(\psi,r)
  &=&
  2\left(1-\beta(r)\vhighs\right)\,\tilde\mu_{40,i}(\psi,r),
  \label{mu22fam}
  \\
  \tilde\mu_{04,i}(\psi,r)
  &=&
  \frac{2}{3}\left[\vhigh
    \left(1-\beta(r)\vhighs\right)\left(2-\beta(r)\vhighs\right)
    - \frac{1}{\beta_\delta}
    \left(\beta_0-\beta(r)\vhighs\right)
    \left(\beta_\infty-\beta(r)\vhighs\right)
  \right]\,\tilde\mu_{40,i}(\psi,r).  
  \label{mu04fam}
\end{eqnarray}
The true velocity moments are connected with the anisotropic velocity
moments through the relation
\begin{equation}
  \tilde{\mu}_{2l,2m,2n}(\psi,r) =
  \frac{1}{\pi} B\left(m+\frac{1}{2},n+\frac{1}{2}\right)\,
  \tilde{\mu}_{2l,2(m+n)}(\psi,r),
\end{equation}
so that
\begin{eqnarray}
  \rho\sigma_r^2(r) 
  &=& 
  \mu_{200}(r) 
  = 
  \sum_{i=1}^N a_{N,i}\,
  \tilde\mu_{20,i}\left(\psi(r),r\vhighs\right),
  \\
  \rho\sigma_\theta^2(r) 
  &=& 
  \mu_{020}(r) 
  = 
  \frac{1}{2}\sum_{i=1}^N a_{N,i}\,
  \tilde\mu_{02,i}\left(\psi(r),r\vhighs\right),
  \\
  \rho\langle v_r^4 \rangle(r) 
  &=& 
  \mu_{400}(r) 
  = \sum_{i=1}^N a_{N,i}\,
  \tilde\mu_{40,i}\left(\psi(r),r\vhighs\right),
\label{mu400}
  \\
  \rho\langle v_\theta^4 \rangle(r) 
  &=& 
  \mu_{040}(r) 
  = \frac{3}{4}\sum_{i=1}^N a_{N,i}\,
  \tilde\mu_{04,i}\left(\psi(r),r\vhighs\right).
\label{mu040}
\end{eqnarray}
Finally, the fourth-order anisotropy profile can be derived by
combining eqs.~(\ref{mu04fam}), (\ref{mu400}) and (\ref{mu040}),
\begin{equation}
  \beta_4(r)
  =
  1 - \frac{\langle v_\theta^4 \rangle}{\langle v_r^4 \rangle}(r)
  =
  \frac{1}{2}\beta(r)\left(3-\beta(r)\vhighs\right) +
  \frac{1}{2\beta_\delta} \left(\beta_0-\beta(r)\vhighs\right)
  \left(\beta_\infty-\beta(r)\vhighs\right).
\end{equation}

\end{document}